# The Strength and Detectability of the YORP Effect in Near-Earth Asteroids: A Statistical Approach

B. Rozitis[a] and S. F. Green[a]

[a]*Planetary and Space Sciences, Department of Physical Sciences, The Open University, Walton Hall, Milton Keynes, MK7 6AA, UK*

No. of Manuscript Pages: 26
No. of Figures: 8
No. of Tables: 3

Please direct editorial correspondence and proofs to:

Benjamin Rozitis
Planetary and Space Sciences
Department of Physical Sciences
The Open University
Walton Hall
Milton Keynes
Buckinghamshire
MK7 6AA
UK

Phone: +44 (0) 1908 655169
Fax: +44 (0) 1908 655667

Email: b.rozitis@open.ac.uk

Email address of co-author: s.f.green@open.ac.uk

## ABSTRACT

In addition to collisions and gravitational forces, it is now becoming widely acknowledged that photon recoil forces and torques from the asymmetric reflection and thermal re-radiation of sunlight are primary mechanisms that govern the rotational evolution of an asteroid. The Yarkovsky-O'Keefe-Radzievskii-Paddack (YORP) effect causes changes in the rotation rate and pole direction of an irregularly shaped asteroid. We present a simple Monte Carlo method to estimate the range of YORP-rotational-accelerations acting on a near-Earth asteroid (NEA) without knowledge of its detailed shape, and to estimate its detectability using light-curve observations. The method requires knowledge of an asteroid's orbital properties and size, and assumes that the future observational circumstances of an asteroid have already been thought through. It is verified by application to the observational circumstances of the seven YORP-investigated asteroids, and is then applied to 540 NEAs with NEOWISE and/or other diameter measurements, and to all NEAs using MPCORB absolute magnitudes. The YORP-detectability is found to be a strong function of the combined asteroid orbital and diameter properties, and is independent of the rotation period for NEAs that don't have very fast or slow rotation rates. The median and 1-sigma spread of YORP-rotational-acceleration expected to be acting on a particular NEA ($d\omega/dt$ in rad yr$^{-2}$) can be estimated from its semimajor axis ($a$ in AU), eccentricity ($e$), and diameter ($D$ in km) by using $|d\omega/dt| = 1.20^{+1.66}_{-0.86} \times 10^{-2} \left( a^2 \sqrt{1-e^2} D^2 \right)^{-1}$ and/or by using $|d\omega/dt| = 1.00^{+3.07}_{-0.81} \times 10^{-2} \left( a^2 \sqrt{1-e^2} D^2 \right)^{-1}$ if the diameter is instead estimated from the absolute magnitude by assuming a geometric albedo of 0.1. The length of a light-curve observational campaign required to achieve a 50 per cent probability of detecting the YORP effect in a particular NEA ($T_{\mathrm{CAM\_50}}$ in yr) can be estimated by using $T_{\mathrm{CAM\_50}} = 12.5 \left( a^2 \sqrt{1-e^2} D^2 \right)^{1/2}$ and/or by using $T_{\mathrm{CAM\_50}} = 13.7 \left( a^2 \sqrt{1-e^2} D^2 \right)^{1/2}$ for an absolute-magnitude-estimated diameter. To achieve a 95 per cent YORP-detection probability, these last two relations need to be multiplied by factors of ~3.4 and ~4.5 respectively. This method and approximate relations will be useful for astronomers who plan to look for YORP-rotational-acceleration in specific NEAs, and for all-sky surveys that may serendipitously observe NEA light-curves.

**Running head:**
The YORP Effect in NEAs

# 1. INTRODUCTION

## 1.1 The YORP Effect

The asteroidal Yarkovsky-O'Keefe-Radzievskii-Paddack (YORP) effect is a change in rotation rate and pole direction caused by the asymmetric reflection and thermal re-radiation of sunlight from an irregularly shaped asteroid (Rubincam 2000). It is related to the Yarkovsky effect, which causes orbital drift, and both have a number of important implications in asteroid science (see review by Bottke et al. 2006). Depending on the asteroid shape asymmetry the YORP-torque can either: continuously increase the rotation rate (spin-up) until the sign is reversed by a re-shaping process encountered at the spin-barrier, or continuously decrease the rotation rate (spin-down) until the asteroid enters a tumbling rotation state. It also has the tendency to shift the pole direction of an asteroid to an asymptotic state that is perpendicular to its orbital plane. The timescale at which this occurs (YORP-timescale) can be shorter than the typical dynamical lifetime of an asteroid, especially for asteroids that are smaller than 40 km in size which are more susceptible to it, and can therefore easily affect its physical and dynamical evolution.

Several signatures characteristic of the YORP effect have been observed in the asteroid population. For example, YORP spin-up and spin-down can explain the observed excess of very fast and slow rotators in asteroids smaller than 40 km in size (Pravec et al. 2008). Small rubble pile asteroids (gravitational bound aggregates) are able to be spun-up so fast that they are forced to change shape and/or undergo mass shedding (Holsapple 2010), and some of their observed shapes reflect this shaping process. Approximately 15 per cent of near-Earth asteroids are inferred to be binaries (Pravec & Harris 2007), and continued YORP spin-up has been demonstrated by numerical simulations to be a possible formation mechanism (Walsh, Richardson & Michel 2008). It has also been recently suggested that YORP-induced rotational fission of contact-binary asteroids is a viable formation mechanism of unbound asteroid pairs (Pravec et al. 2010). Finally, the YORP-induced pole direction changes can explain the clustering of pole directions observed in asteroid families (Vokrouhlický, Nesvorný & Bottke 2003).

Direct detection of YORP-rotational-acceleration has been achieved on near-Earth asteroids (54509) YORP (Lowry et al. 2007; Taylor et al. 2007), (1862) Apollo (Kaasalainen et al. 2007; Ďurech et al. 2008a), and (1620) Geographos (Ďurech et al. 2008b) by observing phase shifts in their rotational light-curves over several years. A fourth probable detection exists for near-Earth asteroid (3103) Eger (Ďurech et al. 2012) which remains to be conclusively confirmed. All have been observed to have an increasing rotation rate and, at present, no asteroid has been observed to have a decreasing rotation rate. YORP effect investigations have also been performed for asteroids (1865) Cerberus, (2100) Ra-Shalom, and (25143) Itokawa but these studies did not detect any changes in rotation rate (Ďurech et al. 2008a, 2012). Table 1 details the physical properties and the YORP effect investigations for these seven asteroids.

## 1.2 YORP Effect Modelling

Many theoretical models of the YORP effect utilising available asteroid shape models have been developed to make predictions, confirm detections, and to assess its impact on the asteroid population (see Table 1 in Rozitis & Green 2012a). In these models, the photon recoil forces and torques from reflected and thermally emitted photons from surface elements of the shape model are calculated, summed across the surface, and time-averaged over the asteroid rotation and orbit to give the overall YORP-torque. The computed YORP-torque is then combined with an estimate of the asteroid's moment of inertia to give the overall spin rate and pole direction changes. The methodology used to calculate the photon torques varies between the models, and typical physical effects that are often included are projected shadows, delayed thermal re-radiation of sunlight caused by non-zero thermal inertia, and Lambertian scattering/emission. Studies utilising these models demonstrate that the YORP-rotational-acceleration is simply inversely proportional to the square of the asteroid diameter and also to the square of the heliocentric distance (Rubincam 2000). It is independent of albedo and thermal inertia (Čapek & Vokrouhlický 2004), and is highly sensitive in a complicated way to the obliquity (Vokrouhlický & Čapek 2002), shape resolution (Breiter et al. 2009), unresolved surface features and roughness (Statler 2009; Rozitis & Green 2012a), and internal bulk density distribution (Scheeres & Gaskell 2008). They also demonstrate that it is equally likely for an asteroid to be spun-up or spun-down by YORP torques without any prior knowledge of its shape.

For the four YORP-detected asteroids, the theoretical predictions using their optical- and radar-derived shape models match the sign and strength of the observed values reasonably well using plausible properties that have been inferred by other observational methods. However, light-curve observations of asteroid (25143) Itokawa fail to see a strong YORP-rotational-deceleration (Ďurech et al. 2008a) that is predicted by YORP effect modelling using the Hayabusa-derived shape models (e.g. Scheeres et al. 2007; Breiter et al. 2009). It remains uncertain as to whether this is caused by a product of specific model assumptions and simplifications, or whether it is a case of simply not knowing the shape to a sufficient resolution or not knowing that there is a non-uniform internal bulk density distribution. For the other two non-detected asteroids, their theoretical models predict changes in rotation rate that are slightly greater than their light-curve detection limits but they have a large degree of uncertainty.

Without knowledge of an asteroid's shape it might seem impracticable to make a prediction of the YORP-rotational-acceleration acting on it. However, Pravec et al. (2008) utilise the median YORP-timescale (i.e. the time it takes for the rotation rate to be doubled/halted) calculated for the sample of asteroid shapes studied by Čapek & Vokrouhlický (2004) to determine a typical magnitude of the YORP-rotational-acceleration acting on main-belt asteroids. Utilising this value they were able to recreate the uniform spin rate distribution of small main-belt asteroids with spin rates between 1 and 9.5 rev $d^{-1}$. They explain an excess of slow rotators, i.e. with spin rates <1 rev $d^{-1}$, as being the result of a weakened YORP effect caused by tumbling rotation states. Rossi, Marzari & Scheeres (2009) compute the effects of YORP on the near-Earth asteroid spin rate distribution using a Monte Carlo numerical model. In their model, the YORP-rotational-acceleration of a particular asteroid is described by a non-dimensional 'YORP-coefficient' that is multiplied by a modified solar constant which is then scaled accordingly to the asteroid's size, density, and orbital properties. This YORP-coefficient contains combined information on the asteroid's shape, moment of inertia, obliquity, and other properties that drive the YORP effect, and its value is chosen from a probability distribution. Model near-Earth asteroid populations were generated with different YORP-coefficient distributions and then dynamically evolved to see which best match the observed spin rate distribution. They found that including YORP is a necessity to match the observations because a model population with no YORP effect deviates significantly from the observed one. However, they were unable to determine a unique YORP-coefficient distribution that best matches the observations because different distributions when run under different simulation circumstances produced very similar results.

### 1.3 Estimating The YORP Effect Detectability

Predictions for directly detecting YORP-rotational-accelerations on near-Earth asteroids other than Itokawa have been made (e.g. Vokrouhlický & Čapek 2002; Čapek & Vokrouhlický 2004; Scheeres 2007). These predictions are generally limited to a small sample of asteroids with optical- or radar-derived shape models, and because of the apparent extreme sensitivity of the YORP effect to small-scale shape variations the results may not be definitive. Kwiatkowski (2010) uses the same approach as Pravec et al. (2008) to estimate the magnitude and the detectability of the YORP-rotational-acceleration acting on a selection of very small near-Earth asteroids observed by the SALT telescope. These studies show that light-curve observations spanning several years to several decades are required to achieve a sufficiently small rotation period uncertainty to resolve any YORP-induced rotation period changes, which has been verified by the observational campaigns used to make the detections for the four YORP-detected asteroids.

In a slightly different but related area of work, Nugent et al. (2012a) has used diameter measurements made by NEOWISE or by other means in the published literature, in combination with an analytical Yarkovsky effect model (Vokrouhlický, Milani & Chesley 2000), to estimate the orbital drifts for 540 near-Earth asteroids. These observations place strong constraints on asteroid diameters and albedos, which are parameters that strongly influence the strength of the Yarkovsky effect. Since thermal conductivity and bulk density are unknown but important parameters they use Monte Carlo modelling to explore how variations in these parameters contribute to uncertainties of the calculated orbital drift. Using their results they identify prediction trends and the twelve most suitable asteroids to look for orbital drift in future observations.

Inspired by the works of Rossi, Marzari & Scheeres (2009), Kwiatkowski (2010), and Nugent et al. (2012a), a simple Monte Carlo method to estimate the strength and detectability of the YORP-rotational-acceleration of an asteroid without knowing its shape is developed in Section 2. The method utilises either a direct diameter measurement of the asteroid or the asteroid's absolute magnitude value; hereafter referred to as diameter-based and absolute-magnitude-based method variants/predictions respectively. The new Monte Carlo method is verified against the observational circumstances for the seven YORP-investigated asteroids in Section 3.1. It is then applied to the 540 near-Earth asteroids with diameter determinations in Section 3.2, and to all currently known near-Earth asteroids utilising their absolute magnitude values in Section 3.3. Further discussion of the results is given in Section 4, and the key results and conclusions are summarised in Section 5.

## 2. STATISTICAL YORP EFFECT MODELLING

### 2.1 Parameterisation Of The YORP Effect Strength

Like Rossi, Marzari & Scheeres (2009), the YORP-rotational-acceleration acting on an asteroid, $d\omega/dt$, given in rad s$^{-2}$ is calculated by

$$\frac{d\omega}{dt} = \frac{G_1}{a^2 \sqrt{1-e^2} \rho D^2} C_Y , \qquad (1)$$

where $G_1$ is a modified solar constant (~6.4 x10$^{10}$ kg km s$^{-2}$), $a$ is the asteroid orbital semimajor axis in km, $e$ is the orbital eccentricity, $\rho$ is the bulk density in kg m$^{-3}$, $D$ is the diameter in m, and $C_Y$ is the non-dimensional YORP-coefficient as described in Section 1.2. This is a suitable parameterisation for calculating the YORP-rotational-acceleration on a general asteroid since it produces a linear change in rotation rate for a constant YORP-coefficient, which is what has been observed for the four YORP-detected asteroids. The YORP-coefficient can be considered constant during an observing campaign because the observing time required for a YORP effect detection is much shorter than the time for the YORP-induced obliquity change to significantly alter the YORP-coefficient. However, such a parameterisation may not be suitable for an asteroid that is either in a tumbling rotation state or is undergoing fast-spin shape deformation, as the YORP-coefficient may change during an observing campaign.

Furthermore, the time it takes for the asteroid rotation rate to be doubled/halted (i.e. the YORP-timescale), $T_{\text{YORP}}$, is given by

$$T_{\text{YORP}} = \omega \left/ \left| \frac{\mathrm{d}\omega}{\mathrm{d}t} \right| \right. , \tag{2}$$

where $\omega$ is the asteroid angular rotation rate. The rate of change in rotation period, $\mathrm{d}P/\mathrm{d}t$, can also be calculated using

$$\frac{\mathrm{d}P}{\mathrm{d}t} = -\frac{P^2}{2\pi} \frac{\mathrm{d}\omega}{\mathrm{d}t} , \tag{3}$$

where $P$ is the asteroid rotation period.

## 2.2 Detecting The YORP Effect

When planning a light-curve observational campaign intending to detect the YORP effect on a particular asteroid then the future observational circumstances of that asteroid play a key role. For example, an asteroid will require a sufficient number of observation data sets within the required time frame. This may be difficult to achieve if the asteroid is consistently in the wrong place to be observed, or is consistently not bright enough because it either is too small, has too low an albedo, is too far away, or has some combination of the previous three. The asteroid will also require an obvious light-curve variation for light-curve folding and period determination, which might not be resolvable for near spherical asteroids or those with close to pole-on configurations at the time of observations. However, most of these observing aspects can already be planned ahead using existing ephemeris computation tools (such as JPL Horizons), and what remains unknown is the length of time to observe an asteroid in order to achieve a detection. At present, there are two ways of extracting YORP effect information from light-curve data: by "direct" comparison of two or more rotation period measurements (e.g. Lowry et al. 2007), and by "fitting" an initial rotation period and a rate of rotation period change (e.g. Kaasalainen et al. 2007; Ďurech et al. 2008a,b, 2012). In either method, the length of observational time required is an important factor in making a YORP effect detection.

In this work, we assume that there are sufficient observational opportunities for a specific asteroid to achieve the required YORP-detection as predicted by the model. If there are insufficient opportunities then this analysis may not be appropriate. To determine whether an asteroid's YORP-rotational-acceleration is detectable in a light-curve observational campaign we consider the "direct" YORP effect extraction method, which is chosen because it can be described by simple equations, and because the length of observing time calculated should be very similar to that required by the "fitting" extraction method. Using the "direct" extraction method therefore requires consideration of the rotation period uncertainties produced using light-curve inversion techniques (e.g. Kaasalainen, Torppa & Muinonen 2001). For a light-curve set that contains more than one apparition and spans a time of $T_{\text{OBS}}$, the rotation period spectrum derived from light-curve inversion contains densely packed $\chi^2$-minima that are separated by an amount $\Delta P$ given by

$$\Delta P = \frac{P^2}{2T_{\text{OBS}}}. \tag{4}$$

This relation derives from the fact that if $P$ is changed by $\Delta P$ then the minima and maxima of the model light-curve will undergo a phase shift of $\pi$. If the best $\chi^2$-minimum of the rotation period spectrum is clearly lower than the others then an uncertainty estimate of a tenth to a hundredth part of $\Delta P$ can be made since the edge of the best $\chi^2$-minimum is very steep (Kaasalainen, Torppa & Muinonen 2001; Torppa et al. 2003). If the neighbouring $\chi^2$-minima are not clearly higher than the best one then the local uncertainty estimate cannot be applied. This may happen when there are only two well-observed apparitions and the exact number of asteroid revolutions between them is uncertain. The more apparitions that are available then the more pronounced the correct $\chi^2$-minimum is. Therefore, a simple way to assess whether an asteroid's YORP-rotational-acceleration is detectable after a certain length of time is to consider a light-curve observational campaign that contains enough apparitions to resolve any ambiguity on the correct $\chi^2$-minimum and spans a total time of $T_{\text{CAM}}$. This campaign can be split into two sets of observations, each with an equal length of time of $T_{\text{OBS}} = T_{\text{CAM}}/2$, where two independent measurements of the rotation period can be made. The uncertainty, $\sigma_{\text{P}}$, of each rotation period measurement is then given by

$$\sigma_{\text{P}} = C_{\text{P}} \frac{P^2}{2T_{\text{OBS}}}, \tag{5}$$

where $C_{\text{P}}$ is a rotation period accuracy coefficient which has a value between 0.01 and 0.1. The absolute change in rotation period between the two campaign measurements, $|\Delta P|$, is given by

$$|\Delta P| = \frac{T_{\text{CAM}}}{2} \left| \frac{dP}{dt} \right|. \tag{6}$$

If a YORP effect detection is to be made at the $X$-sigma confidence level then the absolute change in rotation period must be greater than $X$ times the combined uncertainty of the two rotation period measurements, which gives the detectability condition

$$|\Delta P| \geq 2X\sigma_{\text{P}}. \tag{7}$$

By combining equations (3), (5), (6), and (7) then this detectability condition can also be written as

$$\left| \frac{d\omega}{dt} \right| \geq \frac{8X\pi C_{\text{P}}}{T_{\text{CAM}}^2}, \tag{8}$$

which gives the surprising result that the detectability is independent of rotation period. However, this independence of rotation period may not be applicable to asteroids with very fast and slow rotation rates since they bring additional observing challenges. For a very fast rotating asteroid, very short exposures are required to adequately sample its light-curve for rotation period determination, which may not be possible if the asteroid is insufficiently bright. For a very slow rotating asteroid, at least half coverage of its light-curve is required to obtain an estimate of its rotation period, which would be difficult to obtain during one night on Earth if the rotation period is greater than 24 hours.

## 2.3 Assessing The YORP Effect Dectection Probability

The overall probability of an asteroid to be YORP-detected can be assessed by: generating a series of clones with a range of YORP-rotational-accelerations that are consistent with its physical parameters, and then measuring the fraction of clones that satisfy the detectability condition. A series of clones can be generated via Monte Carlo selection of the physical parameters that drive the YORP effect given in equation (1). Since the orbital properties of asteroids are generally well known then the orbital semimajor axis and eccentricity can be kept fixed at their nominal values for each clone.

Asteroid diameter measurements have typical uncertainties that range from ~1 per cent for in-situ spacecraft observations to ~10 per cent for thermal-IR observations. As in Nugent et al. (2012a), this diameter uncertainty is introduced within the clones by Gaussian distributed noise where the nominal value and uncertainty are the mean and 1-sigma width, respectively, of the standard distribution used (i.e. the diameter-based method variant).

For an asteroid without a measured diameter, its diameter can be estimated from its absolute visual magnitude $H_V$ by

$$D = \frac{10^{-H_V/5} 1329}{\sqrt{p_V}} \, \text{km} \,,$$ (9)

where $p_V$ is the asteroid geometric visual albedo (Fowler & Chillemi 1992). Since the geometric albedo can lie anywhere between 0.02 and 0.7 the absolute magnitude does not usually provide a strong constraint on the asteroid diameter. However, recent observations by NEOWISE have allowed determination of the geometric albedo distribution of near-Earth asteroids (Mainzer et al. 2011), and that the frequency distribution, $f(p_V)$, can be described by a double Gaussian:

$$f\left(p_V\right) = v_0 \exp\left(-\frac{\left(p_V - v_1\right)^2}{2v_2^2}\right) + v_3 \exp\left(-\frac{\left(p_V - v_4\right)^2}{2v_5^2}\right),$$ (10)

where $v_0 = 12.63$, $v_1 = 0.034$, $v_2 = 0.014$, $v_3 = 3.99$, $v_4 = 0.151$, and $v_5 = 0.122$. For asteroids with absolute-magnitude-derived diameters, the diameter uncertainty is introduced within the clones by using Gaussian distributed noise on the absolute magnitude value with a 1-sigma width equal to the absolute magnitude uncertainty, and by randomly selecting the geometric albedo from the frequency distribution given by equation (10). Absolute magnitudes are obtained from the MPCORB database (www.minorplanetcenter.org/iau/MPCORB.html) and like Mainzer et al. (2011) we assume that they have an uncertainty of 0.3 magnitudes. The combination of this absolute magnitude uncertainty with the geometric albedo distribution leads to typical uncertainties of several tens of per cent on the asteroid diameter (i.e. the absolute-magnitude-based method variant).

At present, there is very little information on the bulk density distribution of near-Earth asteroids, but it ranges from ~1000 kg m$^{-3}$ for rubble pile asteroids to ~3000 kg m$^{-3}$ for monolithic bodies (Britt et al. 2002). Again, as in Nugent et al. (2012a), the clone bulk densities are randomly chosen from this range using a uniform frequency distribution.

Finally, suitable values of the non-dimensional YORP-coefficient must be chosen, and as summarised in Table 1 it varies from 0.004 to 0.019 for the four YORP-detected asteroids. As mentioned previously, Rossi, Marzari & Scheeres (2009) do not constrain a unique YORP-coefficient distribution, and find that three different distributions produce similar matches to the observed near-Earth asteroid spin rate distribution when run under different simulation circumstances. These include: a uniform distribution with values ranging from -0.025 to 0.025, a wide Gaussian distribution with width of 0.0125, and a narrow Gaussian distribution with width of 0.0083. These distributions attempt to include all possible combinations of the properties that drive the YORP effect, and the range of values appears to be consistent with those determined theoretically from a large sample of shapes representative of near-Earth asteroids (see figure 5d of Rozitis & Green 2012b). To take this degeneracy into account within the Monte Carlo modelling, YORP-coefficient values are drawn from all three distributions. Fig. 1 compares the three possible YORP-coefficient distributions and the values constrained or measured for the seven YORP-investigated asteroids.

In this work, 3000 clones are generated for each asteroid utilising the distributions described above where 1000 clones are drawn from each of the three YORP-coefficient distributions. The absolute YORP-rotational-acceleration and YORP-timescale for each clone is calculated using equations (1) and (2), which then allow a range of predictions to be calculated for each asteroid that can be characterised by the median prediction and the 1-sigma spread of values surrounding it. The 3-sigma detection probabilities for various observational campaign lengths for each asteroid can be found by counting the number of its clones that satisfy the detection criteria given by equation (8). Here, a constant value of 0.025 is assumed for $C_P$ since it best reproduces the period uncertainties quoted for the four YORP-detected asteroids. Finally, the observational campaign lengths needed to achieve 50 and 95 per cent chances of detecting the YORP effect for a particular asteroid, $T_{CAM\_50}$ and $T_{CAM\_95}$, can be calculated by using values of the absolute YORP-rotational-acceleration that are smaller than 50 and 95 per cent of the clone values, $|d\omega/dt|_{50}$ and $|d\omega/dt|_{95}$, in a re-arranged form of equation (8):

$$T_{CAM\_50} = \sqrt{\frac{8X\pi C_P}{\left|\dfrac{d\omega}{dt}\right|_{50}}} \tag{11}$$

$$T_{CAM\_95} = \sqrt{\frac{8X\pi C_P}{\left|\dfrac{d\omega}{dt}\right|_{95}}} . \tag{12}$$

## 3. APPLICATION TO NEAR-EARTH ASTEROIDS

### 3.1 Verification Using The Seven YORP-Investigated Asteroids

To verify the methodology described in Section 2, it is applied to the observational circumstances of the four YORP-detected asteroids listed in Table 1. Diameter-based and absolute-magnitude-based predictions are made for each asteroid so that comparisons of the different method variants can be made with the measured values. Even though shape models exist for these asteroids they are not used to constrain the range of possible YORP-coefficients, and so the results presented here are purely statistical.

Fig. 2 shows example diameter-based YORP-rotational-acceleration values for 10 per cent of (1862) Apollo's clones as functions of diameter, bulk density, and YORP-coefficient. A horizontal reference line is drawn to indicate the minimum detectable YORP-rotational-acceleration, and any clone data point that is above this line is considered detectable and any point that is below the line is considered non-detectable. Data points chosen from the three different YORP-coefficient distributions are represented by different style markers. Although there is a fairly large degree of scatter in the predictions they follow the obvious trends that the YORP-rotational-acceleration is inversely proportional to the bulk density and to the square of the diameter, and that it is directly proportional to the YORP-coefficient. As Fig. 2 demonstrates, a large proportion of the clones lie above the reference line, which indicates that the light-curve observations of (1862) Apollo have a good chance of detecting a YORP effect acting on it.

Table 2 summarises the diameter-based predictions for each YORP-investigated asteroid including breakdowns for the different YORP-coefficient distributions used. As indicated, the magnitudes of the YORP-rotational-acceleration predictions agree very well with those constrained or measured. In particular, for asteroid (54509) YORP the statistical-derived prediction provides a better match to the observed value than the prediction made by Taylor et al. (2007) using the radar-derived shape model. The Taylor et al. (2007) prediction of YORP-rotational-acceleration differs from that observed by factors ranging from 2.6 to 6.2, and here the 1-sigma prediction range differs by factors ranging from 0.6 to 4.5. Predictions made using the uniform YORP-coefficient distributions have the largest YORP-rotational-acceleration values, and those made from the narrow Gaussian distributions have the smallest values. The wide Gaussian and combined distributions are very similar and produce prediction values that lie near the middle of the range produced by the previous two distributions.

For the total light-curve observation times that are listed in Table 1, asteroid (54509) YORP has the highest detection probability of 0.93, five asteroids [i.e. (1620) Geographos, (1862) Apollo, (1865) Cerberus, (2100) Ra-Shalom, and (25143) Itokawa] have moderate detection probabilities that range from 0.50 to 0.71, and asteroid (3103) Eger has the lowest detection probability of 0.34. These results suggest that a YORP-detection for (3103) Eger would be difficult to achieve using its current light-curve observations. The 4 years of light-curve observations that were made of (54509) YORP is similar to the 5.1 years required to achieve a 95-per-cent-chance of detecting a YORP effect acting on it, and explains its high detection probability. For the five asteroids with moderate detection probabilities their light-curve observations are very similar to the 50-per-cent-chance-observation-times calculated for these asteroids. The 25 years worth of observations for (3103) Eger is shorter than the 31 years required for a 50-per-cent-chance of detection, and explains its low detection probability. Combining their detection probabilities produces an expected value of 4.42 (~4) YORP-detections out of a sample of 7 asteroids, which agrees with the actual number of four YORP-detections achieved. These results indicate that the diameter-based method variant is a good way to estimate the YORP-rotational-acceleration acting on an asteroid and the length of observational time needed to observe it. Here, the 95-per-cent-chance-observation-times are ~3.6 times longer on average than the 50-per-cent-chance-observation-times.

Fig. 3 shows example absolute-magnitude-based YORP-rotational-acceleration values for 10 per cent of (1862) Apollo's clones as functions of absolute magnitude and geometric albedo. For the assumed uncertainty of 0.3 magnitudes there is no obvious trend in the clone YORP-rotational-acceleration values with absolute magnitude. However, there is a trend of increasing YORP-rotational-acceleration with geometric albedo, which highlights how important it is to know the geometric albedo of an asteroid in order to constrain accurately its diameter and also the magnitude of the YORP effect acting on it. Table 3 summarises the absolute-magnitude-based predictions for each YORP-detected asteroid for just the combined YORP-coefficient distribution. The prediction value ranges are greater than those produced by the diameter-based predictions because of the much larger diameter uncertainty that is introduced when the geometric albedo is not known. As a consequence of the larger prediction ranges produced by the absolute-magnitude-based method variant, the YORP-detection probabilities are smaller, and the 50 and 95 per-cent-chance-observation-times are longer, than those produced by the diameter-based method variant. This is especially so for high geometric albedo asteroids [i.e. (1862) Apollo, (3103) Eger, and (25143) Itokawa] where the absolute-magnitude-based method variant underestimates their geometric albedo, overestimates their size, and therefore underestimates their YORP-rotational-acceleration. Combining the detection probabilities for the seven asteroids produces an expected value of 2.94 (~3) YORP-detections, which is slightly less than the actual number of four YORP-detections achieved. Despite the lower accuracy compared to the diameter-based method variant, the absolute-magnitude-based method variant can still provide a reasonable estimate of the YORP-rotational-acceleration acting on an asteroid and the length of observational time needed to observe it. In this case, the 95-per-cent-chance-observation times are ~4.4 times longer on average than the 50-per-cent-chance-observation times.

## 3.2 Diameter-Based Predictions

Table 4 of Nugent et al. (2012a) summarises the properties of the 540 near-Earth asteroids that have diameter measurements made by NEOWISE or by other means from the published literature. These data are used in the diameter-based method variant to estimate the YORP-rotational-acceleration and its detectability of these asteroids utilising the combined YORP-coefficient distribution. Two observational campaign lengths are assumed: a short 5 year one that is similar in length to that used for asteroid (54509) YORP (a dedicated campaign to detect the YORP effect), and a long 30 year one that is similar in length to those used for asteroids (1620) Geographos, (1862) Apollo, and (3103) Eger (campaigns that utilised historical light-curve data).

Fig. 4 shows the detection probabilities for all 540 asteroids as functions of their orbital properties, diameter, and combined orbital-diameter properties. As demonstrated by the large scatter in the first panel, there is no obvious trend with orbital properties. As shown in the second panel, a rough trend exists with diameter which becomes a very good trend once the diameter is combined with the orbital information, as shown in the third panel. This indicates that the asteroid diameter dominates over the orbital properties when assessing the detection probability. For a 5 year observation campaign, the detection probability drops from 1 to 0 between asteroid diameters ranging from ~0.1 to ~1 km. Likewise, for a 30 year observation campaign this asteroid diameter range is from ~0.3 to ~3 km. Out of the 540 asteroids, 294 have a non-zero detection probability with a 5 year observation campaign (i.e. those smaller than ~1 km in diameter), and 508 of them have a non-zero detection probability with a 30 year observation campaign (i.e. those smaller than ~3 km in diameter). These numbers drop down to 95 and 386, respectively, when considering only those asteroids with detection probabilities greater than 50 per cent.

Fig. 5 shows the median and 1-sigma spread of YORP-rotational-acceleration, and the 50 and 95 per-cent-chance-observation-times for all 540 asteroids as a function of their combined orbital-diameter properties. As indicated, all three can be described by a power-law very well and the best fits are shown in the figure captions. These power-law fits can be used to obtain quick estimates of the YORP-rotational-acceleration and the observational time needed to detect it for other near-Earth asteroids not included in the current list of 540 asteroids. The 95-per-cent-chance-observation-times are typically ~3.4 times longer than the 50-per-cent-chance-observation-times. Fig. 5 indicates that asteroids smaller than 0.1 km in diameter require an observational campaign of less than one year in order to achieve a 50-per-cent-chance of detecting a YORP-rotational-acceleration acting on them. For the largest asteroids (~20 km in diameter), Fig. 5 shows that observational campaigns lasting many hundreds of years are required.

## 3.3 Absolute-Magnitude-Based Predictions

Absolute magnitudes and orbital properties for all currently known near-Earth asteroids are available in the MPCORB database (~8800 in total). These data are used in the absolute-magnitude-based method variant to estimate the YORP-rotational-acceleration and its detectability for all currently known near-Earth asteroids. Again, the combined YORP-coefficient distribution is used, and the two observational campaign lengths of 5 and 30 years are assumed.

Fig. 6 shows the detection probabilities for all currently known near-Earth asteroids as functions of their orbital properties, absolute magnitude, and combined orbital-diameter properties. As was demonstrated in Fig. 4, there is an obvious trend with combined orbital-diameter properties where the asteroid diameter, which has been estimated from the absolute magnitude value by using equation (9) and by assuming a geometric albedo of 0.1, has the largest influence on the detection probability. For a 5 year observation campaign, the detection probability drops from 1 to 0 between absolute magnitudes ranging from ~26 to ~18, and for a 30 year observation campaign this range is from ~22 to ~14. Of all currently known near-Earth asteroids, 90.1 and 99.8 per cent have non-zero detection probabilities for observational campaign lengths of 5 and 30 years, respectively. These percentages drop to 49.3 and 90.5, respectively, when considering only those asteroids with detection probabilities greater than 50 per cent.

Fig. 7 shows the median and 1-sigma spread of YORP-rotational-acceleration, and the 50 and 95 per-cent-chance-observation-times for all currently known near-Earth asteroids as a function of their combined orbital-diameter properties. Like Fig. 5, all three trends can be fitted very well with a power-law. In this case, the 95-per-cent-chance-observation-times are typically ~4.5 times longer than the 50-per-cent-chance-observation-times. Fig. 7 indicates that asteroids with absolute magnitudes greater than 29 only require an observational campaign of less than one tenth of a year in order to achieve a 50-per-cent-chance of detecting a YORP-rotational-acceleration acting on them. However, it must be borne in mind that observational opportunities for such objects maybe rare because of the need for close proximity to Earth.

Fig. 8 shows a comparison of the absolute-magnitude-based predictions and the diameter-based predictions for the 540 near-Earth asteroids that have diameter determinations. As indicated, the absolute-magnitude-based predictions tend to slightly underestimate the YORP-rotational-acceleration, and overestimate the 50 and 95 per-cent-chance-observation-times by ~40 and ~80 per cent on average, respectively, when compared with the diameter-based predictions. The absolute-magnitude-based predictions also tend to underestimate the detection probability, where 394 asteroids out of 540 have equal or lower detection probabilities. The same effect is observed in the four YORP-detected asteroids discussed in Section 3.1, and is a consequence of the larger diameter uncertainty which produces a larger YORP-rotational-acceleration uncertainty. The pessimistic nature of the absolute-magnitude-based predictions would generally ensure that 50 and 95 per-cent-chance-observation-times longer than necessary are used.

## 4. DISCUSSION

The Monte Carlo method introduced in Section 2 takes into account random errors in the observations used to constrain some of the asteroid physical properties; however, it does not take into account systematic errors. Pravec et al. (2012) obtained estimates of absolute magnitudes for 583 main-belt and near-Earth asteroids observed by the Ondřejov and Table Mountain Observatory and compared them to values given in the MPCORB database. They find that for large asteroids the database absolute magnitudes are relatively good on average; however, they also find that the database values are systematically too bright for smaller asteroids with absolute magnitudes greater than ~10. The systematic offset of the database values reaches a maximum at an absolute magnitude of ~14 where the mean offset (i.e. $H_{MPCORB}$ - $H$) is -0.4 to -0.5. They find that NEOWISE diameter estimates of main-belt (Masiero et al. 2011) and near-Earth (Mainzer et al. 2011) asteroids were generally stable because diameter estimates resulting from thermal modelling are relatively insensitive to absolute magnitude uncertainties. However, the geometric albedos from these investigations were systematically overestimated because of the systematic underestimate of the database absolute magnitudes. These systematic effects have been confirmed by Williams (2012) through a re-calibration of all asteroid magnitudes reported to the Minor Planet Centre. In terms of significance for this work, the generally stable nature of the NEOWISE diameter estimates means that the diameter-based detection probabilities presented in Section 3.2 should be generally stable too. However, the systematic offset in the database absolute magnitudes means that the absolute-magnitude-based detection probabilities presented in Section 3.3 are being underestimated slightly, which is evident in Fig. 8. This is because the negative offset in the database absolute magnitudes causes an overestimate of the diameter and an underestimate of the YORP-rotational-acceleration of a particular asteroid. Without adjusting for this systematic offset, the absolute-magnitude-based detection probabilities presented in Section 3.3 should be considered as lower bounds. Updated absolute magnitudes for the entire MPCORB database are currently being produced (Gareth Williams personal communication), which should be used in future work to produce updated absolute-magnitude-based predictions.

At present, there are few dedicated long-term programmes for observing the light-curves of many near-Earth asteroids, and good-YORP-effect-candidate and/or observationally-convenient asteroids are observed on an individual basis instead (e.g. Kwiatkowski 2010). However, all-sky surveys designed for other means could be used to obtain asteroid light-curves suitable for rotation period determination and shape modelling. This has already been demonstrated with SuperWASP observations (Parley et al. 2005), and is also expected from future Pan-STARRS (Ďurech et al. 2005) and Gaia (Mignard et al. 2007) observations. Recent advances in light-curve modelling demonstrate that asteroid rotation periods and shapes can be determined from sparse data (Ďurech et al. 2009) and from multiple data sources (Kaasalainen & Viikinkoski 2012). In the near future, YORP-detections could perhaps be made for many near-Earth asteroids using such data.

Obtaining many YORP effect detections would help to constrain the YORP-coefficient and bulk density distributions, and any trends the asteroid physical properties may have with one another. At present, the Monte Carlo method chooses the values of unknown properties independently from other property values whether known or unknown. A YORP-coefficient relation with rotation period might be expected if high YORP-coefficient values produced the fastest and slowest rotating asteroids of a given size. However, some of the fastest rotating asteroids that have been observed with radar exhibit spheroidal 'spinning-top' shapes that have low YORP-coefficient values [e.g. asteroid (66391) 1999 KW4 as observed by Ostro et al. (2006)], and so such a relation might not be that simple. Also, a YORP-coefficient relation might exist with diameter for the smallest asteroids, as recent 3D heat diffusion modelling of the YORP effect has demonstrated that YORP-rotational-acceleration becomes weaker if a monolithic asteroid with high thermal inertia is comparable in size to its diurnal thermal skin depth (Breiter, Vokrouhlický & Nesvorný 2010). A relation seems to exist between rotation period, size, and bulk density, as many of the fastest rotating asteroids greater than ~0.15 km in diameter do not exceed a critical spin rate, which suggests that they are low bulk density rubble piles held together by self-gravitation only (Pravec, Harris & Michalowski 2002). A few asteroids smaller than ~0.15 km in diameter have been observed to be rotating at spin rates much greater than this critical spin rate, e.g. asteroid (54509) YORP, which suggests that they are monolithic bodies with a relatively high bulk density. However, recent estimates of the strength of cohesion forces between grains inside a rubble pile asteroid suggest that they can be strong enough to hold the asteroid together if it rotates faster than the critical spin rate (Scheeres et al. 2010), which might complicate this relation. The bulk density for a particular asteroid could be estimated from its spectral type; for example, C-type asteroids are observed to have lower bulk densities than S-types (Britt et al.

2002). It could also be estimated from Yarkovsky orbital drift measurements that are combined with a suitable Yarkovsky effect model. The bulk density of asteroid (6489) Golevka was determined this way (Chesley et al. 2003) and Yarkovsky orbital drift measurements have been recently made for 54 near-Earth asteroids (Nugent et al. 2012b).

## 5. SUMMARY AND CONCLUSIONS

A simple Monte Carlo method has been developed to estimate the range of possible YORP-rotational-accelerations acting on a near-Earth asteroid and its potential detectability using light-curve observations. The method requires no detailed knowledge of an asteroid's shape, as the YORP-rotational-acceleration is described in terms of the non-dimensional YORP-coefficient, which is randomly chosen from specified probability distributions. Only knowledge of the asteroid's orbital properties and size is required, and other unknown physical properties are also randomly chosen from specified probability distributions. The method utilises either direct diameter measurements or the asteroid absolute magnitude, and takes into account their uncertainties. The future observational circumstances of an asteroid of interest, which are also key to achieving a YORP-detection, are assumed to have been already thought through, such that the length of time required by a light-curve observational campaign remains the only unknown variable. It is found that the YORP-rotational-acceleration detectability is independent of rotation period for asteroids that don't have very fast or slow rotation rates.

Both method variants were applied to the observational circumstances of the seven YORP-investigated asteroids, and were verified that they predicted the same magnitude of the YORP-rotational-acceleration as that constrained or measured for each asteroid. For the actual observation campaign lengths used, asteroids (1620) Geographos, (1862) Apollo, (1865) Cerberus, (2100) Ra-Shalom, (3103) Eger, (25143) Itokawa, and (54509) YORP, had YORP-detection probabilities of 0.50, 0.57, 0.71, 0.69, 0.34, 0.68, and 0.93 respectively using the diameter-based method variant. Combining these detection probabilities produces an expected value of 4.42 (~4) YORP-detections, which agrees with the actual number of four YORP-detections achieved. Using the absolute-magnitude-based method variant these YORP-detection probabilities were 0.35, 0.26, 0.56, 0.07, 0.28, and 0.86 respectively, which combine to produce an expected value of 2.94 (~3) YORP-detections. These results also suggest that a YORP-detection for asteroid (3103) Eger would be difficult to achieve using its current light-curve observations.

Applying the diameter-based method variant to 540 near-Earth asteroids with diameter determinations reveal a strong detection probability trend with combined orbital-diameter properties. The detection probability drops from 1 to 0 for asteroid diameters ranging from ~0.1 to ~1 km for a 5 year observation campaign, and for diameters ranging from ~0.3 to ~3 km for a 30 year observation campaign. Out of 540 asteroids, 294 and 508 have non-zero detection probabilities with observation campaigns lasting 5 and 30 years respectively. For a greater than 50-per-cent-chance of detection these numbers are 95 and 386 respectively. The median and 1-sigma spread of YORP-rotational-acceleration (in rad yr$^{-2}$) expected to be acting on a near-Earth asteroid can be estimated from its combined orbital-diameter properties using $\left| d\omega/dt \right| = 1.20^{+1.66}_{-0.86} \times 10^{-2} \left( a^2 \sqrt{1-e^2} D^2 \right)^{-1}$ (where $a$ is in AU and $D$ is in km). Similarly, the 50-per-cent-chance-observation-time (in yr) can be estimated using $T_{\text{CAM\_50}} = 12.5 \left( a^2 \sqrt{1-e^2} D^2 \right)^{1/2}$ with the 95-per-cent-chance-observation-time being ~3.4 times longer.

Applying the absolute-magnitude-based method variant to all currently known near-Earth asteroids also reveals a strong detection probability trend with combined-orbital diameter properties. In this case, the detection probability drops from 1 to 0 for asteroid absolute magnitudes ranging from ~26 to ~18 for a 5 year observation campaign, and for absolute magnitudes ranging from ~22 to ~14 for a 30 year observation campaign. 90.1 and 99.8 per cent of all currently known near-Earth asteroids have non-zero detection probabilities for observation campaigns lasting 5 and 30 years respectively. For a greater than 50-per-cent-chance of detection these percentages are 49.3 and 90.5 respectively. The median and 1-sigma spread of YORP-rotational-acceleration (in rad $yr^{-2}$) expected to be acting on a near-Earth asteroid can be estimated from its combined orbital-diameter properties using $\left|\mathrm{d}\omega/\mathrm{d}t\right| = 1.00^{+3.07}_{-0.81} \times 10^{-2} \left(a^2 \sqrt{1-e^2} D^2\right)^{-1}$ (where $a$ is in AU and $D$ is estimated from the absolute magnitude value assuming a geometric albedo of 0.1 and is given in km). Similarly, the 50-per-cent-chance-observation-time (in yr) can be estimated using $T_{\mathrm{CAM\_50}} = 13.7\left(a^2 \sqrt{1-e^2} D^2\right)^{1/2}$ with the 95-per-cent-chance-observation-time being ~4.5 times longer.

This method and approximate relations will be useful for astronomers who plan to look for YORP-rotational-acceleration in specific near-Earth asteroids, and for all-sky surveys that may serendipitously observe near-Earth asteroid light-curves. Finally, the method could also be adapted to apply to main-belt asteroids or other orbital groups.

## Acknowledgments

We are grateful to the reviewer Dr. Stephen Lowry for several suggested refinements to the manuscript. The work of BR is supported by the UK Science and Technology Facilities Council (STFC).

## Tables

Table 1: Summary of the seven YORP-investigated asteroids.

| Asteroid | (1620) Geographos | (1862) Apollo | (1865) Cerberus | (2100) Ra-Shalom | (3103) Eger | (25143) Itokawa | (54509) YORP |
|---|---|---|---|---|---|---|---|
| Semimajor Axis (AU) | 1.245 | 1.470 | 1.080 | 0.832 | 1.405 | 1.324 | 1.006 |
| Eccentricity | 0.336 | 0.560 | 0.467 | 0.436 | 0.354 | 0.280 | 0.230 |
| Absolute Magnitude | 15.6 | 16.3 | 16.8 | 16.1 | 15.4 | 19.2 | 22.7 |
| Mean Diameter (km) | $2.56 \pm 0.05$ | $1.45 \pm 0.22$ | $1.60 \pm 0.24$ | $2.30 \pm 0.20$ | $1.80 \pm 0.30$ | $0.327 \pm 0.001$ | $0.113 \pm 0.002$ |
| Geometric Albedo | $0.16 \pm 0.01$ | $0.25 \pm 0.08$ | $0.13 \pm 0.04$ | $0.12 \pm 0.02$ | $0.38 \pm 0.13$ | $0.345 \pm 0.002$ | $0.115 \pm 0.004$ |
| Rotation Period (hr) | $5.223336 \pm 0.000002$ | $3.065448 \pm 0.000003$ | $6.80328 \pm 0.00001$ | $19.8201 \pm 0.0004$ | $5.710156 \pm 0.000007$ | $12.1323 \pm 0.0002$ | $0.20289941 \pm 0.00000001$ |
| YORP-Rotational-Acceleration (rad yr$^{-2}$) | $(1.5 \pm 0.2)$ x10$^{-3}$ | $(7.3 \pm 1.6)$ x10$^{-3}$ | $\|d\omega/dt\| < 1.1$ x10$^{-3}$ | $-5.3$ x10$^{-3}$ $< d\omega/dt < 2.7$ x10$^{-3}$ | $(1.9 \pm 0.8)$ x10$^{-3}$ | $\|d\omega/dt\| < 2.0$ x10$^{-2}$ | $0.47 \pm 0.05$ |
| YORP-Timescale (Myr) | $6.9 \pm 0.9$ | $2.5 \pm 0.5$ | $> 7.6$ | $> 0.52$ | $5.2 \pm 2.2$ | $> 0.23$ | $0.58 \pm 0.06$ |
| YORP-Coefficient* | $(1.0 \pm 0.1)$ x10$^{-2}$ | $(1.9 \pm 0.4)$ x10$^{-2}$ | $\|C_Y\| < 2.0$ x10$^{-3}$ | $-1.2$ x10$^{-2}$ $< C_Y < 0.6$ x10$^{-2}$ | $(7.8 \pm 3.4)$ x10$^{-3}$ | $\|C_Y\| < 2.5$ x10$^{-3}$ | $(4.2 \pm 0.4)$ x10$^{-3}$ |
| Light-curve Observations | 1969-2008 | 1980-2007 | 1980-2009 | 1978-2009 | 1987-2012 | 2000-2007 | 2001-2005 |
| Light-curve Total Time Length (yr) | 39 | 27 | 29 | 31 | 25 | 7 | 4 |
| YORP Light-curve Reference | Ďurech et al. (2008b) | Ďurech et al. (2008a) | Ďurech et al. (2012) | Ďurech et al. (2012) | Ďurech et al. (2012) | Ďurech et al. (2008a) | Lowry et al. (2007) |
| Size Reference | Hudson & Ostro (1999) | Harris (1998) | Mainzer et al. (2011) | Shepard et al. (2008) | Mainzer et al. (2011) | Fujiwara et al. (2006) | Taylor et al. (2007) |

*Calculated using nominal diameter and by assuming a bulk density of 2000 kg m$^{-3}$.

Table 2: Summary of diameter-based YORP effect predictions for the seven YORP-investigated asteroids.

| | Asteroid | (1620) Geographos | (1862) Apollo | (1865) Cerberus | (2100) Ra-Shalom | (3103) Eger | (25143) Itokawa | (54509) YORP |
|---|---|---|---|---|---|---|---|---|
| **Uniform $C_Y$ Distribution** | YORP-Rotational-Acceleration (rad yr$^{-2}$) | $(1.8^{+1.6}/_{-1.3})$ x10$^{-3}$ | $(4.6^{+5.2}/_{-3.2})$ x10$^{-3}$ | $(6.5^{+7.4}/_{-4.5})$ x10$^{-3}$ | $(5.3^{+5.1}/_{-3.7})$ x10$^{-3}$ | $(2.9^{+3.4}/_{-2.0})$ x10$^{-3}$ | $(9.8^{+8.4}/_{-7.1})$ x10$^{-2}$ | $1.4^{+1.2}/_{-1.0}$ |
| | YORP-Timescale (Myr) | $5.8^{+14.2}/_{-2.8}$ | $3.9^{+9.0}/_{-2.1}$ | $1.2^{+2.9}/_{-0.7}$ | $0.5^{+1.2}/_{-0.3}$ | $3.4^{+7.9}/_{-1.8}$ | $0.05^{+0.12}/_{-0.02}$ | $0.20^{+0.48}/_{-0.09}$ |
| | Detection Probability | 0.67 | 0.71 | 0.82 | 0.81 | 0.48 | 0.79 | 0.95 |
| | 50% Observation Time (yr) | 32 | 20 | 17 | 19 | 26 | 4 | 1.2 |
| | 95% Observation Time (yr) | 110 | 77 | 65 | 69 | 98 | 15 | 4.1 |
| **Wide Gaussian $C_Y$ Distribution** | YORP-Rotational-Acceleration (rad yr$^{-2}$) | $(1.2^{+1.5}/_{-0.9})$ x10$^{-3}$ | $(3.1^{+4.5}/_{-2.2})$ x10$^{-3}$ | $(4.4^{+6.4}/_{-3.1})$ x10$^{-3}$ | $(3.6^{+4.8}/_{-2.5})$ x10$^{-3}$ | $(2.0^{+2.9}/_{-1.4})$ x10$^{-3}$ | $(6.6^{+8.0}/_{-4.6})$ x10$^{-2}$ | $0.94^{+1.16}/_{-0.67}$ |
| | YORP-Timescale (Myr) | $8.6^{+21.3}/_{-4.7}$ | $5.8^{+13.8}/_{-3.5}$ | $1.8^{+4.3}/_{-1.1}$ | $0.8^{+1.7}/_{-0.4}$ | $5.0^{+12.1}/_{-2.9}$ | $0.07^{+0.17}/_{-0.04}$ | $0.29^{+0.72}/_{-0.16}$ |
| | Detection Probability | 0.50 | 0.58 | 0.73 | 0.71 | 0.33 | 0.70 | 0.94 |
| | 50% Observation Time (yr) | 39 | 25 | 21 | 23 | 31 | 5 | 1.4 |
| | 95% Observation Time (yr) | 120 | 78 | 65 | 72 | 99 | 17 | 4.4 |
| **Narrow Gaussian $C_Y$ Distribution** | YORP-Rotational-Acceleration (rad yr$^{-2}$) | $(0.80^{+1.10}/_{-0.58})$ x10$^{-3}$ | $(2.1^{+3.1}/_{-1.5})$ x10$^{-3}$ | $(3.0^{+4.4}/_{-2.2})$ x10$^{-3}$ | $(2.4^{+3.2}/_{-1.7})$ x10$^{-3}$ | $(1.3^{+2.0}/_{-1.0})$ x10$^{-3}$ | $(4.3^{+5.7}/_{-3.1})$ x10$^{-2}$ | $0.61^{+0.83}/_{-0.44}$ |
| | YORP-Timescale (Myr) | $13^{+35}/_{-8}$ | $8.5^{+22.8}/_{-5.1}$ | $2.7^{+7.2}/_{-1.6}$ | $1.2^{+3.1}/_{-0.7}$ | $7.2^{+19.6}/_{-4.3}$ | $0.11^{+0.27}/_{-0.06}$ | $0.45^{+1.17}/_{-0.26}$ |
| | Detection Probability | 0.32 | 0.41 | 0.59 | 0.56 | 0.20 | 0.54 | 0.89 |
| | 50% Observation Time (yr) | 49 | 30 | 25 | 28 | 38 | 7 | 1.8 |
| | 95% Observation Time (yr) | 170 | 103 | 86 | 96 | 130 | 23 | 6.1 |
| **Combined $C_Y$ Distribution** | YORP-Rotational-Acceleration (rad yr$^{-2}$) | $(1.2^{+1.5}/_{-0.9})$ x10$^{-3}$ | $(3.2^{+4.5}/_{-2.3})$ x10$^{-3}$ | $(4.5^{+6.4}/_{-3.3})$ x10$^{-3}$ | $(3.6^{+4.8}/_{-2.6})$ x10$^{-3}$ | $(2.0^{+2.9}/_{-1.5})$ x10$^{-3}$ | $(6.5^{+8.1}/_{-4.8})$ x10$^{-2}$ | $0.94^{+1.16}/_{-0.68}$ |
| | YORP-Timescale (Myr) | $8.6^{+23.0}/_{-4.8}$ | $5.7^{+16.0}/_{-3.5}$ | $1.8^{+5.0}/_{-1.1}$ | $0.8^{+2.1}/_{-0.5}$ | $4.8^{+14.0}/_{-2.9}$ | $0.07^{+0.18}/_{-0.04}$ | $0.29^{+0.78}/_{-0.16}$ |
| | Detection Probability | 0.50 | 0.57 | 0.71 | 0.69 | 0.34 | 0.68 | 0.93 |
| | 50% Observation Time (yr) | 39 | 24 | 21 | 23 | 31 | 5 | 1.4 |
| | 95% Observation Time (yr) | 140 | 88 | 73 | 82 | 110 | 19 | 5.1 |

Table 3: Summary of absolute-magnitude-based YORP effect predictions for the seven YORP-investigated asteroids.

| | Asteroids | (1620) Geographos | (1862) Apollo | (1865) Cerberus | (2100) Ra-Shalom | (3103) Eger | (25143) Itokawa | (54509) YORP |
|---|---|---|---|---|---|---|---|---|
| **Combined $C_Y$ Distribution** | Absolute-Magnitude-Derived Diameter (km) | $3.1^{+2.7}/_{-1.1}$ | $2.3^{+2.0}/_{-0.8}$ | $1.8^{+1.6}/_{-0.6}$ | $2.4^{+2.2}/_{-0.9}$ | $3.4^{+3.0}/_{-1.2}$ | $0.58^{+0.52}/_{-0.21}$ | $0.12^{+0.10}/_{-0.04}$ |
| | YORP-Rotational-Acceleration (rad yr$^{-2}$) | $(0.67^{+2.13}/_{-0.54})$ x10$^{-3}$ | $(1.0^{+3.2}/_{-0.8})$ x10$^{-3}$ | $(2.9^{+9.1}/_{-2.3})$ x10$^{-3}$ | $(2.5^{+7.9}/_{-2.0})$ x10$^{-3}$ | $(0.43^{+1.37}/_{-0.35})$ x10$^{-3}$ | $(1.6^{+5.1}/_{-1.3})$ x10$^{-2}$ | $0.69^{+2.18}/_{-0.55}$ |
| | YORP-Timescale (Myr) | $16^{+64}/_{-12}$ | $18^{+74}/_{-14}$ | $3^{+12}/_{-2}$ | $1.1^{+4.6}/_{-0.9}$ | $22^{+91}/_{-17}$ | $0.28^{+1.16}/_{-0.21}$ | $0.39^{+1.62}/_{-0.30}$ |
| | Detection Probability | 0.35 | 0.26 | 0.56 | 0.56 | 0.07 | 0.28 | 0.86 |
| | 50% Observation Time (yr) | 53 | 43 | 26 | 27 | 66 | 11 | 1.7 |
| | 95% Observation Time (yr) | 240 | 190 | 110 | 120 | 290 | 48 | 7.4 |

**Figure Captions**

Figure 1: Profiles representing the different types of YORP-coefficient distribution used. The dashed, solid, and dotted lines correspond to the uniform, wide Gaussian, and narrow Gaussian distributions as defined by Rossi, Marzari & Scheeres (2009) respectively. The markers indicate the YORP-coefficient values that have been constrained or measured for the seven YORP-investigated asteroids for comparison.

Figure 2: Diameter-based YORP-rotational-acceleration values as a function of diameter, bulk density, and YORP-coefficient for 10 per cent of (1862) Apollo's clones. The different style markers represent values determined by the different YORP-coefficient distributions as indicated by the legend in the top left panel. The solid horizontal lines represent the minimum detectable YORP-rotational-acceleration for the light-curve observational circumstances of (1862) Apollo.

Figure 3: Absolute-magnitude-based YORP-rotational-acceleration values as a function of absolute magnitude and geometric albedo for 10 per cent of (1862) Apollo's clones. The different style markers represent values determined by the different YORP-coefficient distributions as indicated by the legend in the left panel. The solid horizontal lines represent the minimum detectable YORP-rotational-acceleration for the light-curve observational circumstances of (1862) Apollo.

Figure 4: Diameter-based YORP-rotational-acceleration detection probabilities for 540 near-Earth asteroids as a function of orbital properties, diameter, and combined orbital-diameter properties. The grey and black markers represent detection probabilities for light-curve observational campaigns lasting 5 and 30 years respectively.

Figure 5: Diameter-based median and 1-sigma spread of YORP-rotational-acceleration predictions (left panel), and 50 and 95 per-cent-chance-observation-times (right panel) for 540 near-Earth asteroids as a function of diameter in km. The lines are the best power-law fits to the trends indicated, and have the following equations (where $a$ is in AU and $D$ is in km): $\left| \mathrm{d}\omega/\mathrm{d}t \right| = 1.20^{+1.66}_{-0.86} \times 10^{-2} \left( a^2 \sqrt{1-e^2} D^2 \right)^{-1}$ (median and 1-sigma spread of YORP-rotational-acceleration in rad yr$^{-2}$), $T_{\mathrm{CAM\_50}} = 12.5 \left( a^2 \sqrt{1-e^2} D^2 \right)^{1/2}$ (50-per-cent-chance-observation-time in yr), and $T_{\mathrm{CAM\_95}} = 42.5 \left( a^2 \sqrt{1-e^2} D^2 \right)^{1/2}$ (95-per-cent-chance-observation time in yr).

Figure 6: Absolute-magnitude-based YORP-rotational-acceleration detection probabilities for all currently known near-Earth asteroids as a function of orbital properties, absolute magnitude, and combined orbital-diameter properties. The diameter has been estimated from the absolute magnitude value by using equation (9) and by assuming a geometric albedo of 0.1. The grey and black markers represent detection probabilities for light-curve observational campaigns lasting 5 and 30 years respectively.

Figure 7: Absolute-magnitude-based median and 1-sigma spread of YORP-rotational-acceleration predictions (left panel), and 50 and 95 per-cent-chance-observation-times (right panel) for all currently known near-Earth asteroids as a function of absolute magnitude. The lines are the best power-law fits to the trends indicated, and have the following equations (where $a$ is in AU and $D$ in km): $\left|\mathrm{d}\omega/\mathrm{d}t\right| = 1.00^{+3.07}_{-0.81} \times 10^{-2} \left(a^2 \sqrt{1-e^2} D^2\right)^{-1}$ (median and 1-sigma spread of YORP-rotational-acceleration in rad yr$^{-2}$), $T_{\mathrm{CAM\_50}} = 13.7 \left(a^2 \sqrt{1-e^2} D^2\right)^{1/2}$ (50-per-cent-chance-observation-time in yr), and $T_{\mathrm{CAM\_95}} = 61.3 \left(a^2 \sqrt{1-e^2} D^2\right)^{1/2}$ (95-per-cent-chance-observation time in yr). The diameter has been estimated from the absolute magnitude value by using equation (9) and by assuming a geometric albedo of 0.1.

Figure 8: Comparison between diameter-based ($D$) and absolute-magnitude-based ($H$) YORP effect predictions for 540 near-Earth asteroids with diameter determinations. The lines in all panels represent the trends if the two predictions were exactly equal to one another. The grey and black markers in the top right panel represent detection probabilities for light-curve observational campaigns lasting 5 and 30 years respectively.

**Figures**

Figure 1:

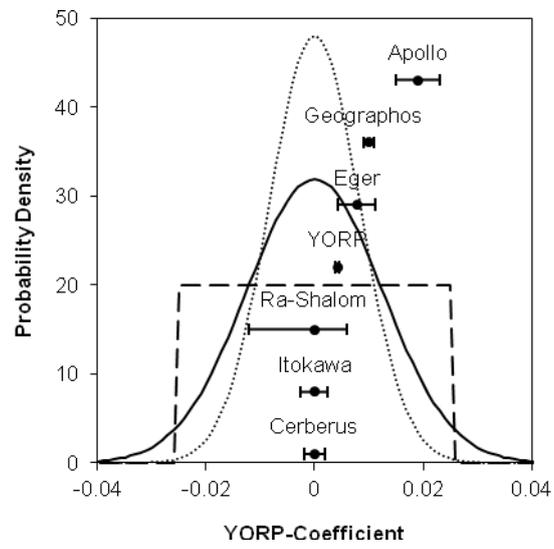

Figure 2:

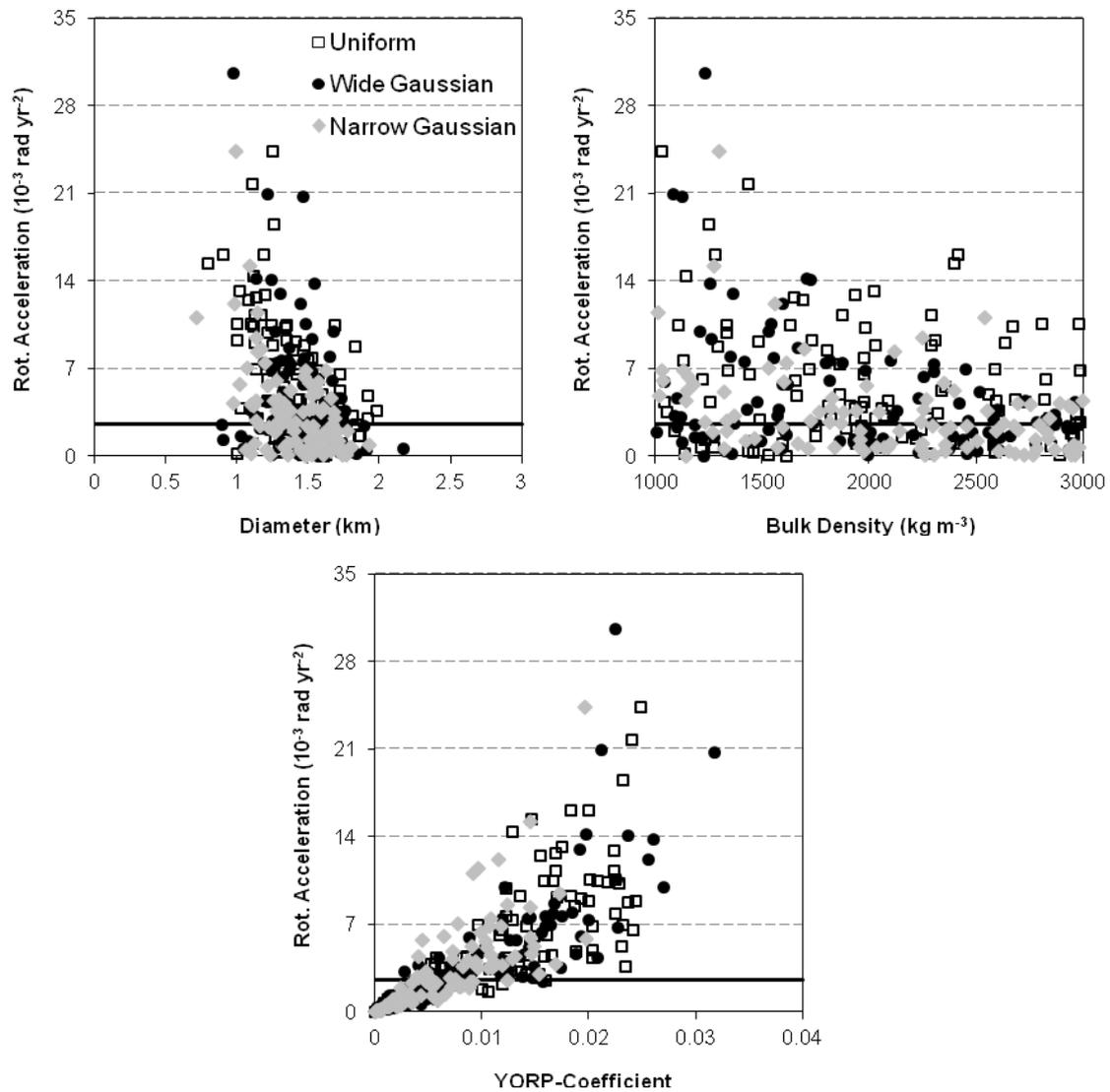

Figure 3:

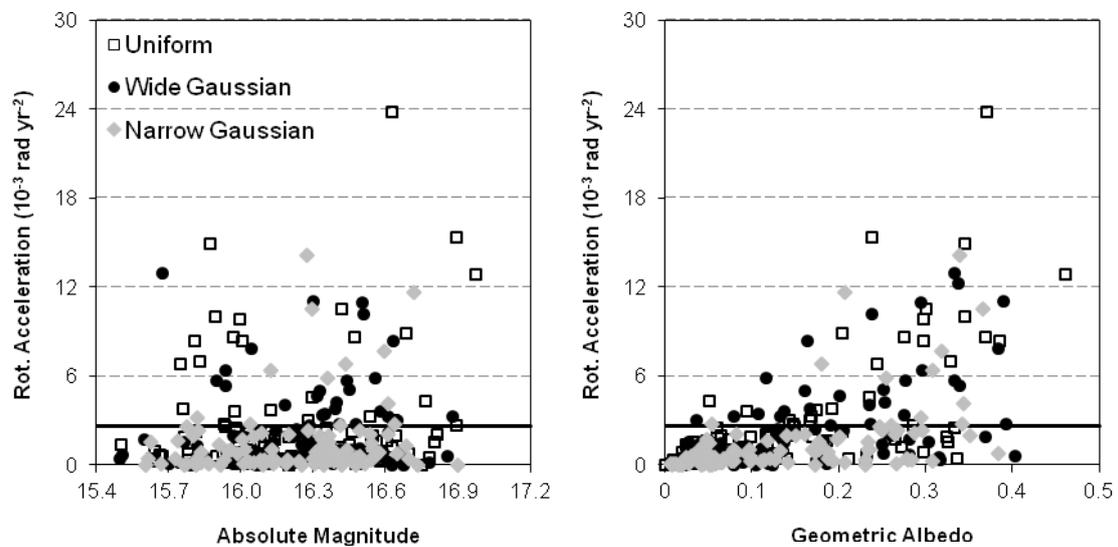

Figure 4:

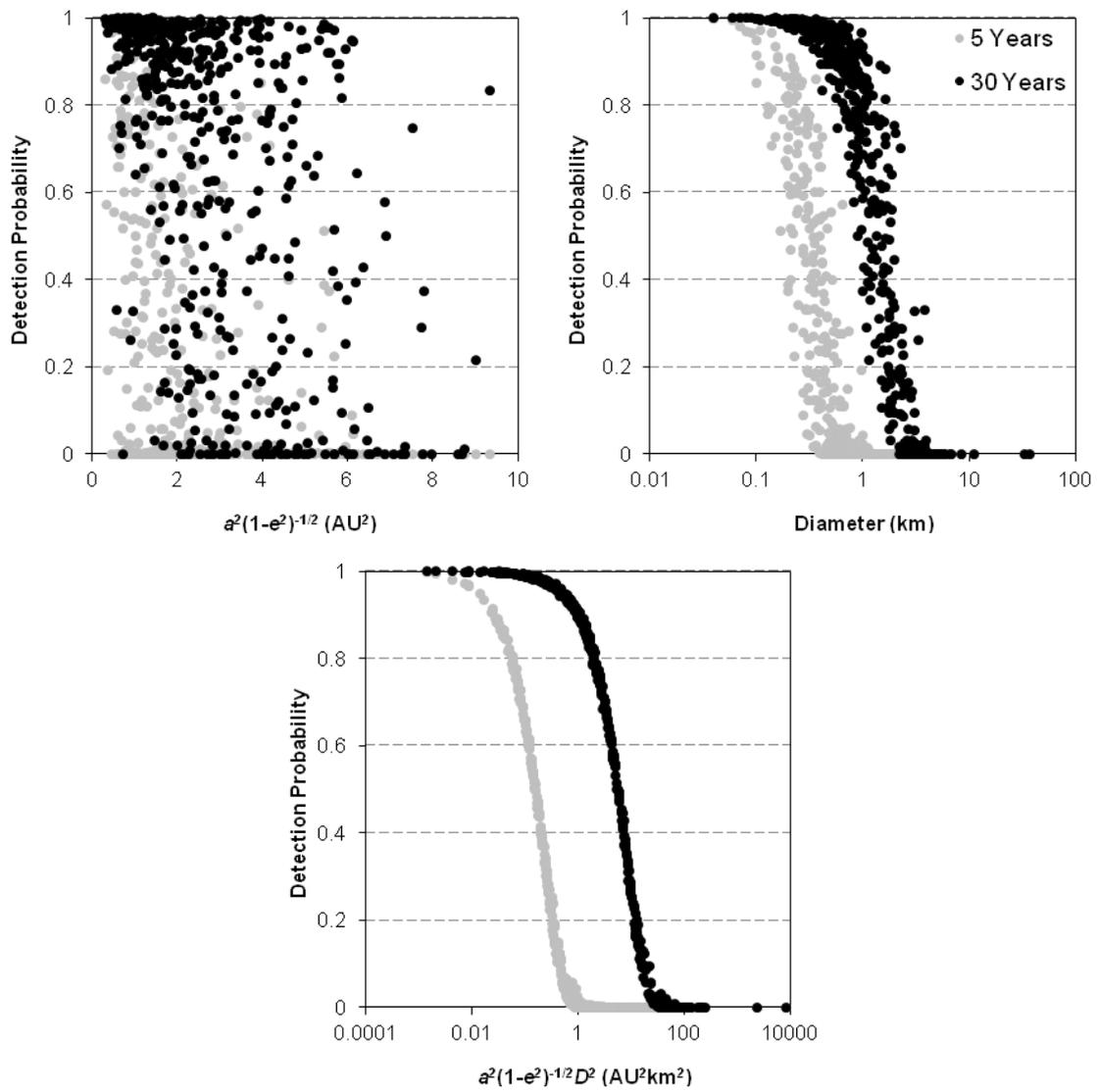

Figure 5:

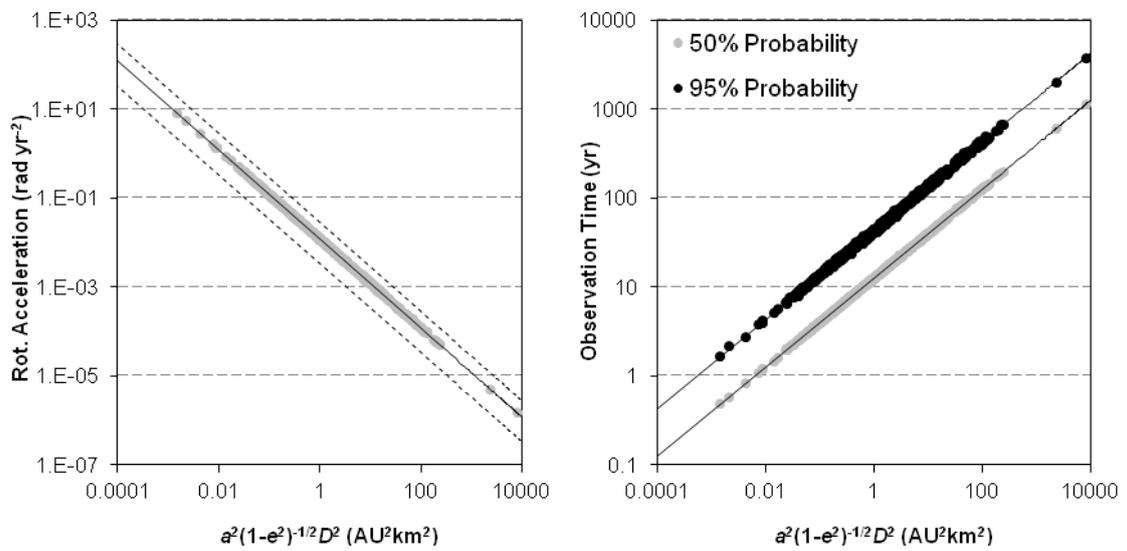

Figure 6:

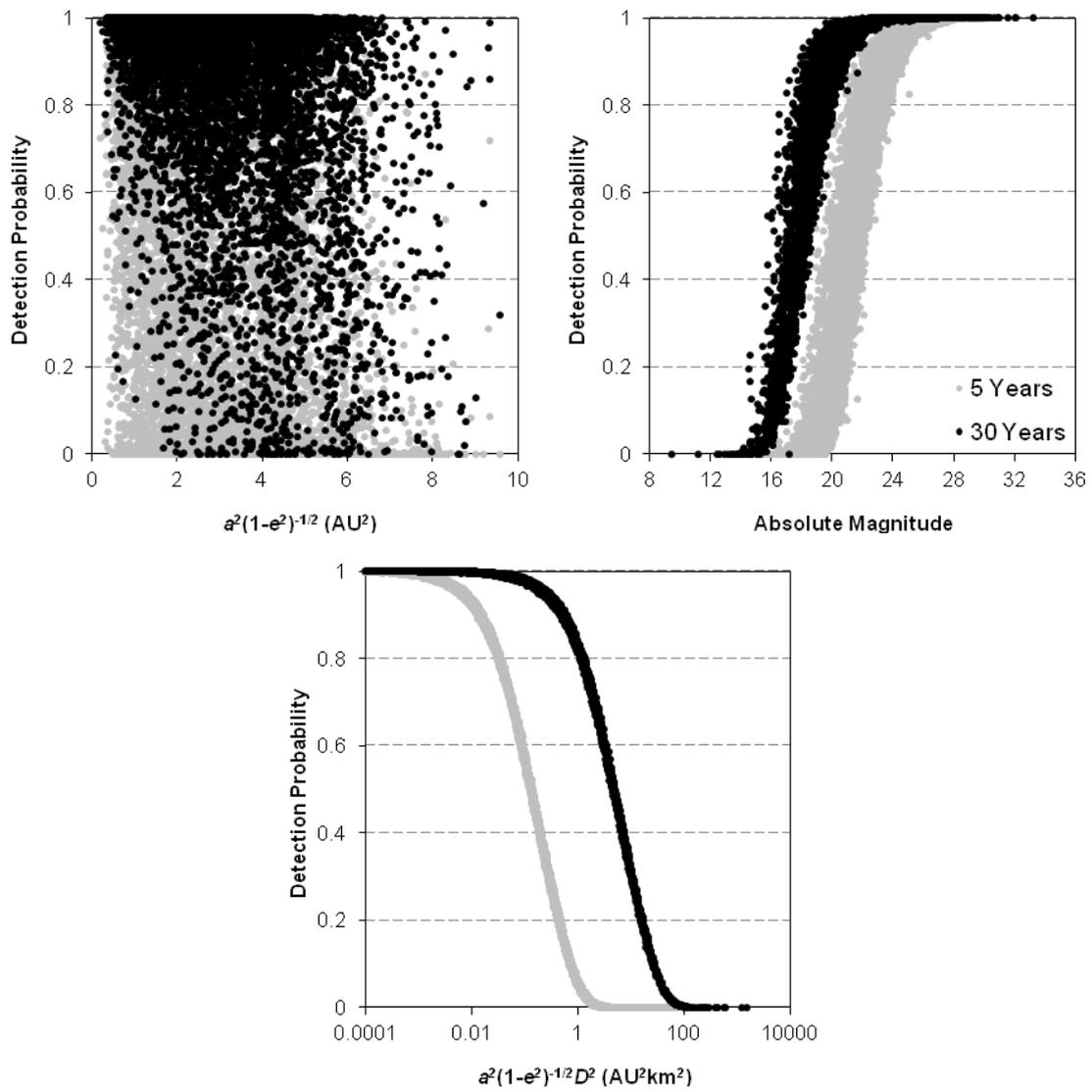

Figure 7:

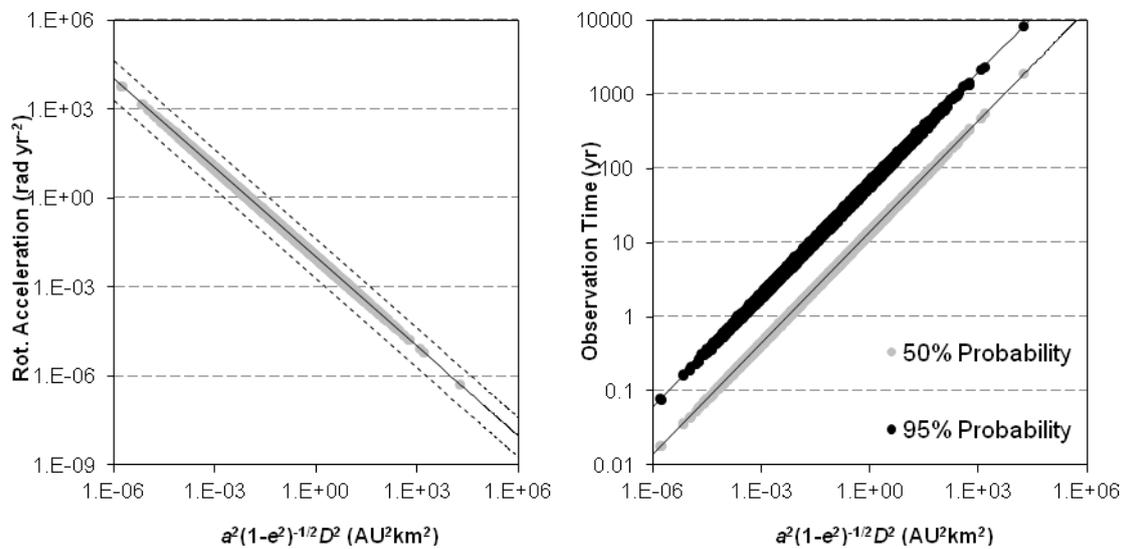

Figure 8:

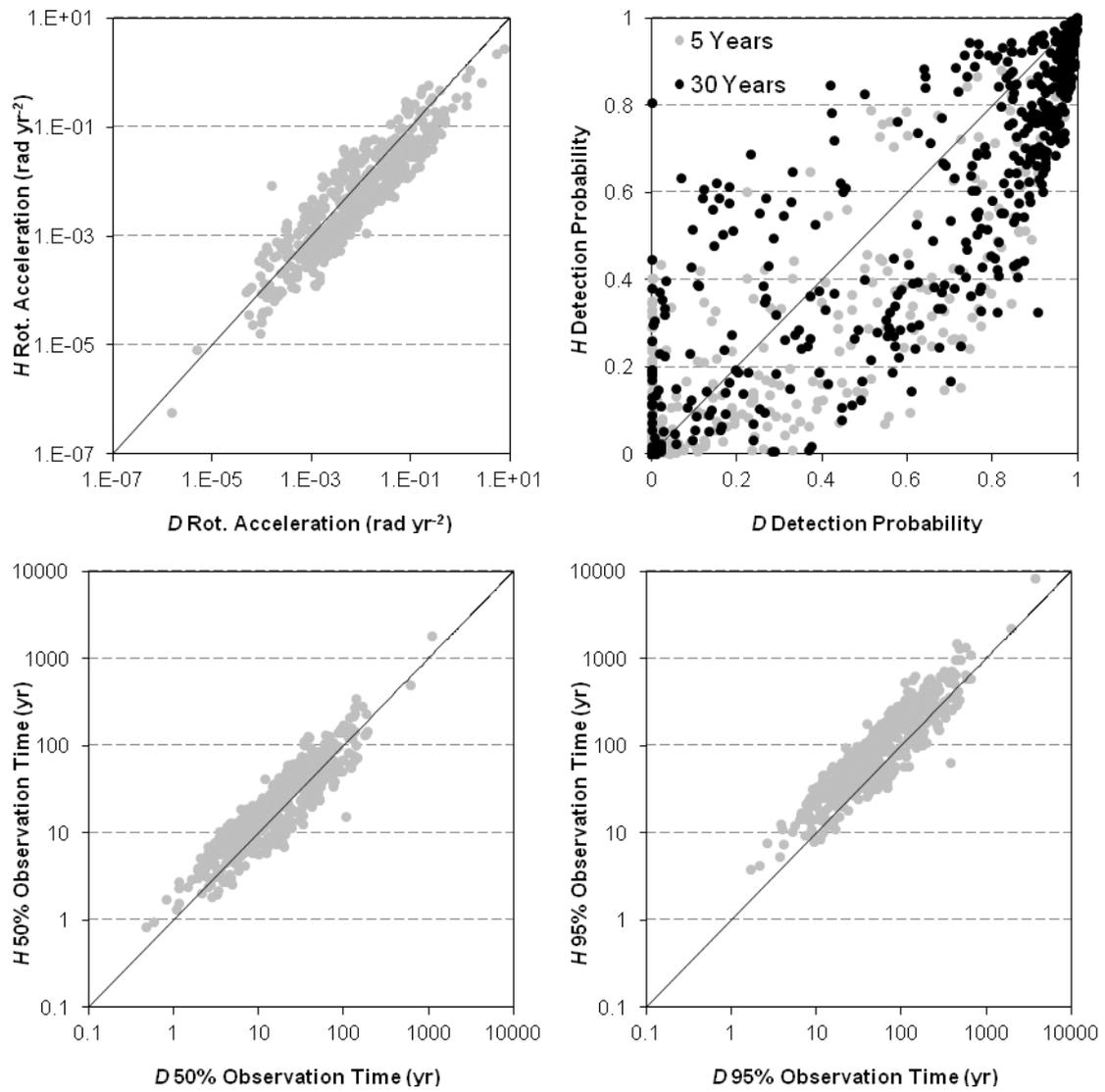